\documentclass[epj]{svjour}
\usepackage{amsmath}
\usepackage{amssymb}
\usepackage{epsfig}
\usepackage{longtable}
\newcommand{\beq}{\begin{equation}}
\newcommand{\eeq}{\end{equation}}
\newcommand{\bea}{\begin{eqnarray}}
\newcommand{\eea}{\end{eqnarray}}

\begin{document}
\title{Screened potential and quarkonia properties at high temperatures}
\author{J. Vijande\inst{1,2}, G. Krein\inst{3}, and A. Valcarce\inst{1}}

\institute{
Departamento de F\'\i sica Fundamental, Universidad de Salamanca, E-37008
Salamanca, Spain \and
Departamento de F\' \i sica At\'omica, Molecular y Nuclear e IFIC,
Universidad de Valencia - CSIC, E-46100 Burjassot, Valencia, Spain
\and
Instituto de F\'{\i}sica Te\'{o}rica, Universidade Estadual
Paulista, Rua Pamplona, 145 - 01405-900 S\~{a}o Paulo, SP, Brazil
}
\date{Received: date / Revised version: date}

\abstract{
We perform a quark model calculation of the quarkonia $b\overline{b}$ and
$c\overline{c}$ spectra using smooth and sudden string breaking potentials.
The screening parameter is scale dependent and can be related to an effective
running gluon mass that has a finite infrared fixed point. A temperature
dependence for the screening mass is motivated by lattice QCD simulations
at finite temperature. Qualitatively different results are obtained for
quarkonia properties close to a critical value of the deconfining
temperature when a smooth or a sudden string breaking potential is used.
In particular, with a sudden string breaking potential quarkonia radii remain
almost independent of the temperature up to the critical point, only well above
the critical point the radii increase significantly. Such a behavior will impact
the phenomenology of quarkonia interactions in medium, in particular for
scattering dissociation processes.}


\authorrunning{J. Vijande et al.}
\titlerunning{Screened potential and quarkonia properties above the
deconfinement transition}

\maketitle

\section{Introduction}
\label{sec:intro}

The study of hot hadronic matter may yield a particularly clear picture
of the physics of quark confinement~\cite{Matsui:1986dk}. At high temperatures,
color screening may be strong enough to lead to the dissolution
of quark-antiquark states~\cite{Moc08}. Because of the large mass of the heavy quarks
in charmonium and bottomonium, the velocity of the heavy quarks is small
enough such that the binding effects in quarkonia at zero temperature might be
understood in terms of a nonrelativistic potential. Color screening could then
be masked in terms of in-medium modification of the interquark forces.

Nature of the confining potential has been a challenge for lattice QCD
studies. Quenched and unquenched lattice QCD calculations were able to
probe the linear raising potential between heavy color sources~\cite{Bali:2000gf}.
It has been also recently numerically investigated
the transition of the static quark-antiquark
string into a static-light meson-antimeson system, in other words string
breaking in QCD~\cite{Bali:2005fu}. These studies drove the idea of screening
of the color forces at zero temperature as a consequence of the polarization
of vacuum. This effect observed numerically should be enhanced in hot
hadronic matter with observable consequences.

Recent publications~\cite{Gonzalez:2003gx,Vijande:2003gk} have emphasized
the use of a screened potential in place of one linearly rising
with inter-quark distance in a quark-model description of the hadronic
spectrum. In these publications properties of hea\-vy quarkonia ($c\bar c$
and $b\bar b$ bound states) such as masses, spin-spin splittings, leptonic
widths and radiative decays have been calculated from a simple screened
funnel quark potential model. Although the quality of the calculated
spectra~\cite{Gonzalez:2003gx} is similar to that of other quark model
calculations~\cite{Eichten:1979ms} that employ nonscreened confinement,
important differences arise. The most salient ones are the finite number of
quark-antiquark bound states and the pattern of energy differences of the
higher excited states. The finiteness of the bound-state spectrum has
interesting implications in the light-quark sector. In particular,
the predicted~\cite{Vijande:2003gk} number of states is in almost perfect
agreement with the experimentally observed states, a fact that might shed
new light on the so-called missing resonance problem -- for a recent review,
see Ref.~\cite{Capstick:2000qj}.

Screening of the potential is due to quark-antiquark creation from the
vacuum as the interquark distance is increased and leads to the
breaking of the color string that would be formed in the absence of sea
quarks~\cite{Bali:2000gf}. Such a string breaking has been confirmed in
lattice QCD calculations~\cite{Bali:2005fu}. It has also been suggested
by the observation of nonlinear hadronic Regge trajectories~\cite{Gold00}.
A quite rapid crossover from a linear rising to a flat potential is
well established in SU(2) Yang-Mills theories~\cite{Yan00}.
In addition, the mentioned recent lattice QCD calculations
have also shown that the breaking is quite sudden, the 
$Q\bar Q$ potential saturates sharply for a breaking distance
of the order of 1.25 fm corresponding to a saturation 
energy of about twice the $B-$meson ($Q\bar q$) mass, indicating
that the formation of two heavy-light subsystems ($B,\bar B$)
is energetically favored. 
This information has been implemented in a quark model scheme~\cite{Swa06} showing that,
as a consequence of coupled channels above the physical
thresholds (corresponding to the opening of decay channels),
the description becomes progressively less accurate
high in the spectrum. Moreover, the mixing with the continuum
can also modify the short-range part of the interaction.
Nonetheless, an effective (renormalized) nonscreened
potential continues being useful up to energies not too far
above the lowest physical threshold.
At sufficiently high baryon densities and/or temperatures, one would expect
that such a screening would be even stronger.

However, recent results of
lattice QCD simulations of charmonium correlation functions at finite temperature
have shown some rather unexpected results~\cite{Umeda:2002vr,Asakawa:2003re,Datta:2003ww}.
The spectral functions in some channels display narrow peaks at temperatures $T$
well above the deconfinement temperature~$T_c$. Peaks in these mesonic correlation
functions indicate that the quark and the antiquark are strongly correlated,
leading to the interpretation that bound states of the heavy charm quarks can
possibly survive above the deconfinement temperature. The results came to a
surprise since early expectations~\cite{Matsui:1986dk} were that $c\bar c$ bound
states (like the $J/\Psi$) would dissolve already at temperatures close to~$T_c$.

These lattice results have stirred renewed interest~
\cite{{Shuryak:2003ty},{Shuryak:2004tx},{Wong:2004zr},{Alberico:2005xw},{Mannarelli:2005pa},{Mocsy:2005qw}}
in incorporating finite temperature effects in a potential model. There is a long history~\cite{Satz:2005hx}
on the use of temperature-dependent potentials in a Schr\"odinger equation to study quarkonium properties
at finite temperatures and the tlattice results have brought new physical insight into the problem.
Recent studies have attempted to incorporate this insight into the phenomenology
of the modified potentials. Within these approaches a temperature dependence for the potential
is extracted from lattice results for the finite temperature free energy of a
static quark-antiquark pair. A problem with such an strategy is that the
free-energy is not itself a potential energy since it contains an entropy
contribution. One consequence of such a parametrization of the binding potential
is that entropy smooths out any sudden breaking of the string. This in turn has
the effect that quarkonia properties close to the critical temperature have a
smooth temperature dependence.

\begin{figure}[t]
\epsfig{file=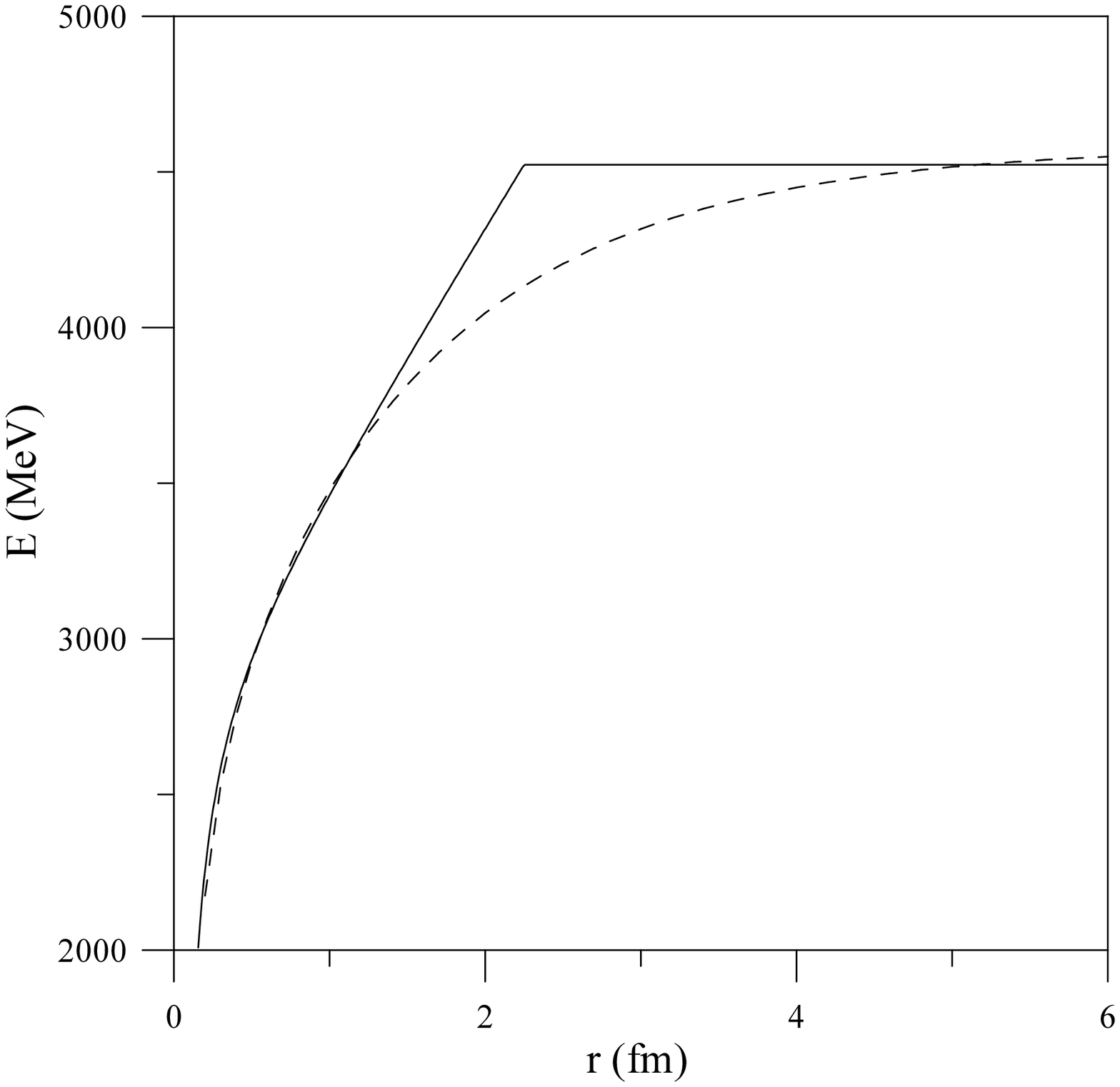,height=70mm}
\epsfig{file=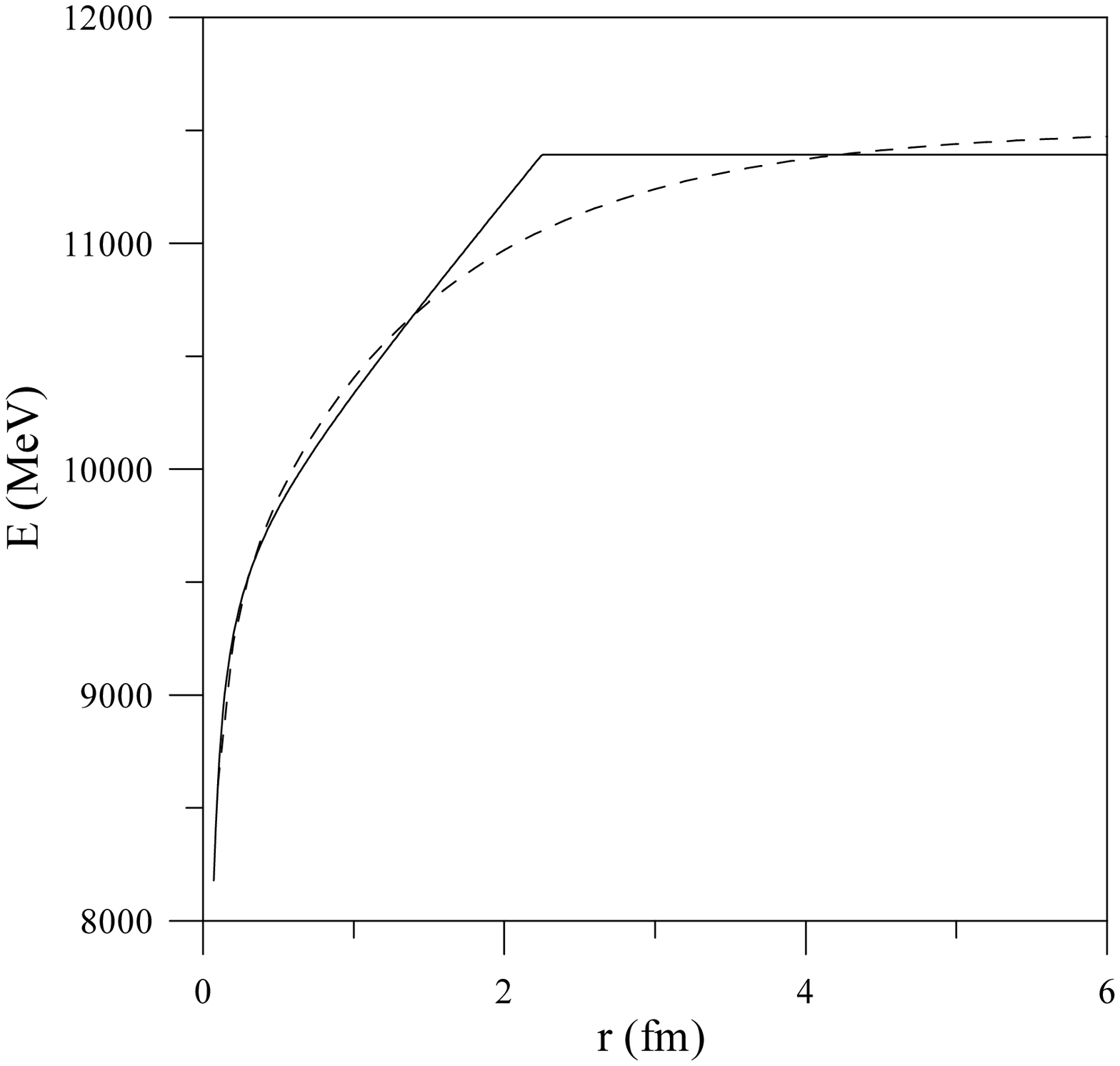,height=70mm}
\caption{Smooth and sudden potentials for charmonium (upper panel) and
bottomonium (lower panel) pseudoscalar $S$ wave states.
In all cases the value of the corresponding
constituent quark masses (2$m_c$ or 2$m_b$) has been added.}
\label{fig:pot-cc}
\end{figure}

Although at the moment such a finite temperature quark model, contrary
to the zero temperature case, cannot be justified as some limiting approximation
that follows from a systematic effective field theory, it does seem to provide
a simple phenomenological attempt to bring insight into the problem. At zero
temperature, potential models for quarkonia can be derived in QCD from first
principles as the leading order approximation of an effective field theory known
as potential nonrelativistic QCD~\cite{{Bali:2000gf},{Brambilla:2004jw}}. A finite
temperature generalization of this approach has been attempted only very
recently~\cite{Laine:2007qy}, but no string
breaking is accessible in such an approach. Work on similar grounds to obtain
a finite temperature potential to be used in a Schr\"odinger equation was done
in Ref.~\cite{Brambilla:2008cx}

The purpose of the present paper is to show that one obtains qualitatively
different behavior of quarkonia properties close to the critical temperature
when a sudden string-breaking potential is used. In particular, the radius
of a quarkonium bound state remains almost independent of the temperature up
to the critical temperature, when starts to increase abruptly.
Such a behavior will impact the
phenomenology of quarkonia interactions in medium, in particular for
scattering dissociation processes
\cite{{Hadjimichef:1998rx},{Wong:1999zb},{Hilbert:2007hc},{Haidenbauer:2007jq}},
a topic very important for the experimental programs of heavy-ion collisions like
the experiments at the FAIR facility at the GSI laboratory in Germany, in particular
for the CBM experiment~\cite{CBM-Fair}.

The paper is organized as follows. In the next section we discuss the
models incorporating smooth and sudden breaking at zero and finite temperature and
we present our numerical results. A model for the temperature dependence of the
screening parameters is discussed in Section ~\ref{sec:temperature}. Conclusions and
Perspectives are summarized in Section~\ref{sec:concl}.

\begin{table}[t]
\caption{$T=0$ parameters.}
\begin{center}
\begin{tabular}{|c|cc|}
\hline
				& Smooth	& Sudden \\
\hline
$\sigma$ (MeV fm$^{-1}$)	& 1470		& 800	\\
$\mu$ (fm$^{-1}$)		& 0.71		& $-$ \\
$1/r_b$ (fm$^{-1}$)		& $-$		& 0.44 \\
$\alpha$ (MeV fm)		& 96		& 106 \\
$r_0$ (fm)			& 0.38		& 0.36 \\
$m_c$ (MeV)			& 1264		& 1385 \\
$m_b$ (MeV)			& 4724		& 4820 \\
\hline
\end{tabular}
\end{center}
\label{tab:T=0}
\end{table}

\section{Smooth versus sudden string breaking}
\label{sec:smoo_sudd}

Our approach in the present paper will be purely phenomenological. Initially,
for the purposes of investigating the consequences of sudden string breaking
at finite temperatures it is not necessary to adopt any specific model for
the temperature dependence of screening parameters. We will calculate the spectrum
of charmonium and bottomonium using two different potentials, one with smooth
string breaking and another with sudden string breaking. We then vary the screening
parameters and investigate its effect over the energies, radii and decay
constants. In the next section we will discuss possible relations of our
phenomenological approach to different models that parametrize the temperature
dependence of the screening parameters. This will allow us to relate changes of
the observables with temperature.

\begin{table}[t]
\caption{$c\overline{c}$ bound state masses (in MeV) with smooth
and sudden string breaking at $T =0$ up to four radial excitations. Experimental
masses taken from the PDG \cite{PDG08}.}
\begin{center}
\begin{tabular}{|ccc|c|}
\hline
State ($nL_{2S+1})$&$M_{smooth}$ 	& $M_{sudden}$ 	& $M_{exp}$ \\
\hline
$1S_1$		& 2979		& 2976		& $2979.8\pm 1.2$ \\
$1S_3$		& 3099		& 3096		& $3096.916\pm 0.011$                \\
$1P_1$		& 3491		& 3453		& $3525.93\pm0.27$ \\
$1P_3$		& 3521		& 3482		& $3493.87$            \\
$2S_1$		& 3639		& 3600		& $3637\pm4$ \\
$2S_3$		& 3686		& 3654		& $3686.093\pm 0.034$                \\
$1D_1$		& 3790		& 3745		& \\
$1D_3$		& 3801		& 3757		& $3772.4\pm 1.1$     \\
$2P_1$		& 3907		& 3897		& \\
$2P_3$		& 3923		& 3917		& $3929\pm 5$\\
$3S_1$		& 4008		& 4028		& \\
$3S_3$		& 4035		& 4067		& $4039\pm 1$        \\
$2D_1$		& 4098		& 4128		& \\
$2D_3$		& 4105		& 4138		& $4153\pm 3$\\
$3P_1$		& 4178		& 4266		& \\
$3P_3$		& 4189		& 4282		& \\
$4S_1$		& 4248		& 4384		& \\
$4S_3$		& 4265		& 4413		& $4421\pm 4$ \\
$3D_1$		& 4307		& 4458		& \\
$3D_3$		& 4312		& 4467		& \\
$4P_1$		& 4361		&	$-$    	& \\
$4P_3$		& 4368		&	$-$    	& \\
\hline
\end{tabular}
\end{center}
\label{tab:ccT=0}
\end{table}

\begin{table}[t]
\caption{$b\overline{b}$ bound state masses (in MeV) with smooth
and sudden string breaking at $T=0$ up to six radial excitations.
Experimental masses taken from the PDG \cite{PDG08}.}
\begin{center}
\begin{tabular}{|ccc|c|}
\hline
State ($nL_{2S+1})$& $M_{smooth}$ 	& $M_{sudden}$ 	& $M_{exp}$ \\
\hline
$1S_1$	& 9434		& 9432		&  \\
$1S_3$	& 9459		& 9463		& $9460.30\pm 0.26$                \\
$1P_1$	& 9951		& 9929		& \\
$1P_3$	& 9958		& 9936		& $9888.1$                  \\
$2S_1$	& 10059		& 10012		& \\
$2S_3$	& 10068		& 10022		& $10023.26\pm 0.31$               \\
$1D_1$	& 10218		& 10167		& \\
$1D_3$	& 10221		& 10171		& $10161.1\pm 1.7$  \\
$2P_1$	& 10320		& 10263		& \\
$2P_3$	& 10324		& 10267		& $10252.2$                 \\
$3S_1$	& 10404		& 10339		& \\
$3S_3$	& 10409		& 10345		& $10355.2\pm 0.5$                 \\
$2D_1$	& 10502		& 10441		& \\
$2D_3$	& 10505 	& 10443		& \\
$3P_1$	& 10583		& 10528		& \\
$3P_3$	& 10586		& 10531		& \\
$4S_1$	& 10651		& 10599		& \\
$4S_3$	& 10655		& 10604		& $10579.4\pm 1.2$                 \\
$3D_1$	& 10721		& 10677		& \\
$3D_3$	& 10722 	& 10678 	& \\
$4P_1$	& 10787		& 10758		& \\
$4P_3$	& 10789		& 10760		&\\
$5S_1$  & 10843         & 10825   &\\
$5S_3$	& 10846		& 10829		& $10865\pm 8$                 \\
$4D_1$  & 10895         & 10889   & \\
$4D_3$  & 10896         & 10891   & \\
$5P_1$  & 10949         & 10966   &  \\
$5P_3$  & 10951         & 10968   &  \\
$6S_3$  & 10998         & 11034   & $11019\pm 8$   \\
\hline
\end{tabular}
\end{center}
\label{tab:bbT=0}
\end{table}

We implement smooth string breaking in the potential as
\begin{equation}
V_{smooth}(r) = \frac{\sigma}{\mu} \left( 1 - e^{- \mu r} \right) + V_{OGE}(r) ,
\label{Vsmooth}
\end{equation}
where $V_{OGE}(r)$ is the one-gluon exchange (OGE) potential given by
\begin{equation}
V_{OGE}(r) = - \frac{\alpha}{r} + \alpha\frac{\hbar^2}{m_q m_{\bar q}}
\frac{ e^{-r/r_0} }{rr^2_0}\left(\vec \sigma_1 \cdot \vec \sigma_2\right).
\end{equation}
Here $\sigma$ and $\alpha$ are phenomenological parameters to be fixed by
fitting the quarkonium spectrum, and the delta function of the OGE spin-spin part has been
smoothed out with a parameter $r_0$. The asymptotic limit $r\to\infty$ of the potential
is a constant, given by
\begin{equation}
V_{smooth}(r \rightarrow \infty) = \frac{\sigma}{\mu}.
\end{equation}
Sudden string breaking is implemented as
\begin{equation}
V_{sudden}(r)  = \left\{
\begin{array}{ll}
\sigma \, r + V_{OGE}(r) , & \hspace{0.25cm}
r < r_b \\[3mm]
\sigma r_b + V_{OGE}(r_b),
& \hspace{0.25cm} r \geq r_b \, .
\end{array}
\right.
\label{Vsudden}
\end{equation}
The asymptotic limiting value of this
potential is spin($S$)-dependent and it is given by
\begin{eqnarray}
V_{sudden}(r \rightarrow\infty) &=& \sigma r_b - \frac{\alpha}{r_b}\nonumber\\
&+&\left[S(S+1)\right]\alpha\frac{\hbar^2}{m_q m_{\bar q}}\frac{e^{-r_b/r_0}}{r_0^2r_b}\, .
\end{eqnarray}

We adjust parameters to obtain a reasonable description of the lowest states.
Although they are effective parameters they are not unphysical, see Ref.~\cite{Gon06}
for a detailed discussion of their value.
Their values are given in Table~\ref{tab:T=0} (we refer to these as the $T=0$ parameters).
In Fig.~\ref{fig:pot-cc} we plot the potentials for the $S$ wave pseudoscalar (spin equal to zero)
states of charmonium and bottomonium. The differences between the smooth and sudden
string-breaking potentials are concentrated in the region of $r \simeq 2$~fm. We have
solved the Schr\"odinger equation for the above potentials for charmonium and bottomonium
using standard Numerov techniques~\cite{Koo90}. Our results, together with the available experimental
values taken from the PDG \cite{PDG08} are shown in Tables~\ref{tab:ccT=0}
and \ref{tab:bbT=0}. Many of the experimental results shown in the tables have no definite
assignment of orbital angular momentum.
Therefore, we have identified the states guided by the results of
our model. As a general trend, one sees that the higher orbital excitations are better
described by a sudden breaking potential. This important result, already observed
in the baryon spectra~\cite{Vijande:2003gk}, imply that these states are
very sensitive to the form of the confining potential and as such they will be an ideal benchmark
to provide clues on the nature of the screening behavior of the potential.

\begin{figure}[t]
\epsfig{file=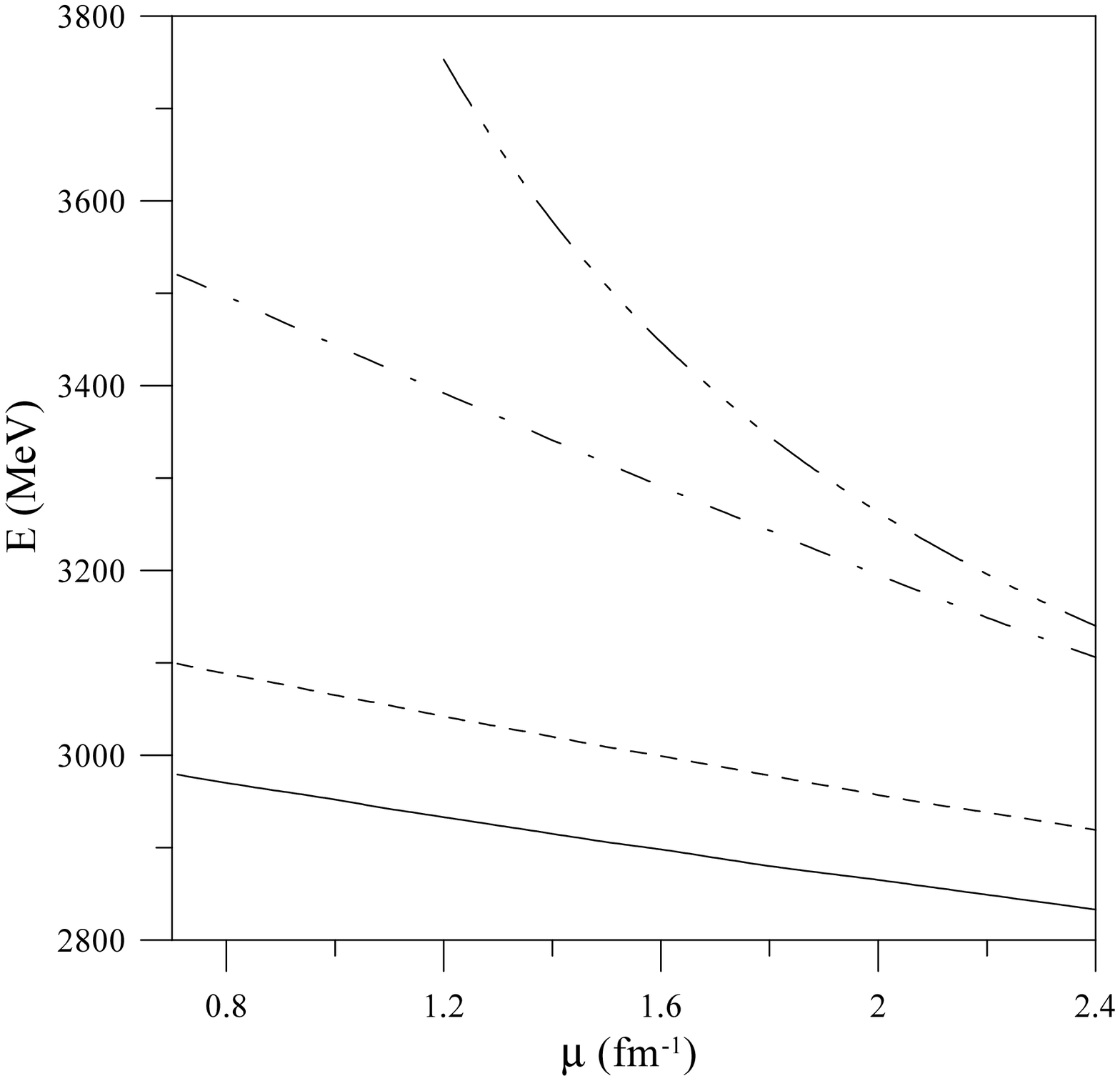,height=70mm}
\epsfig{file=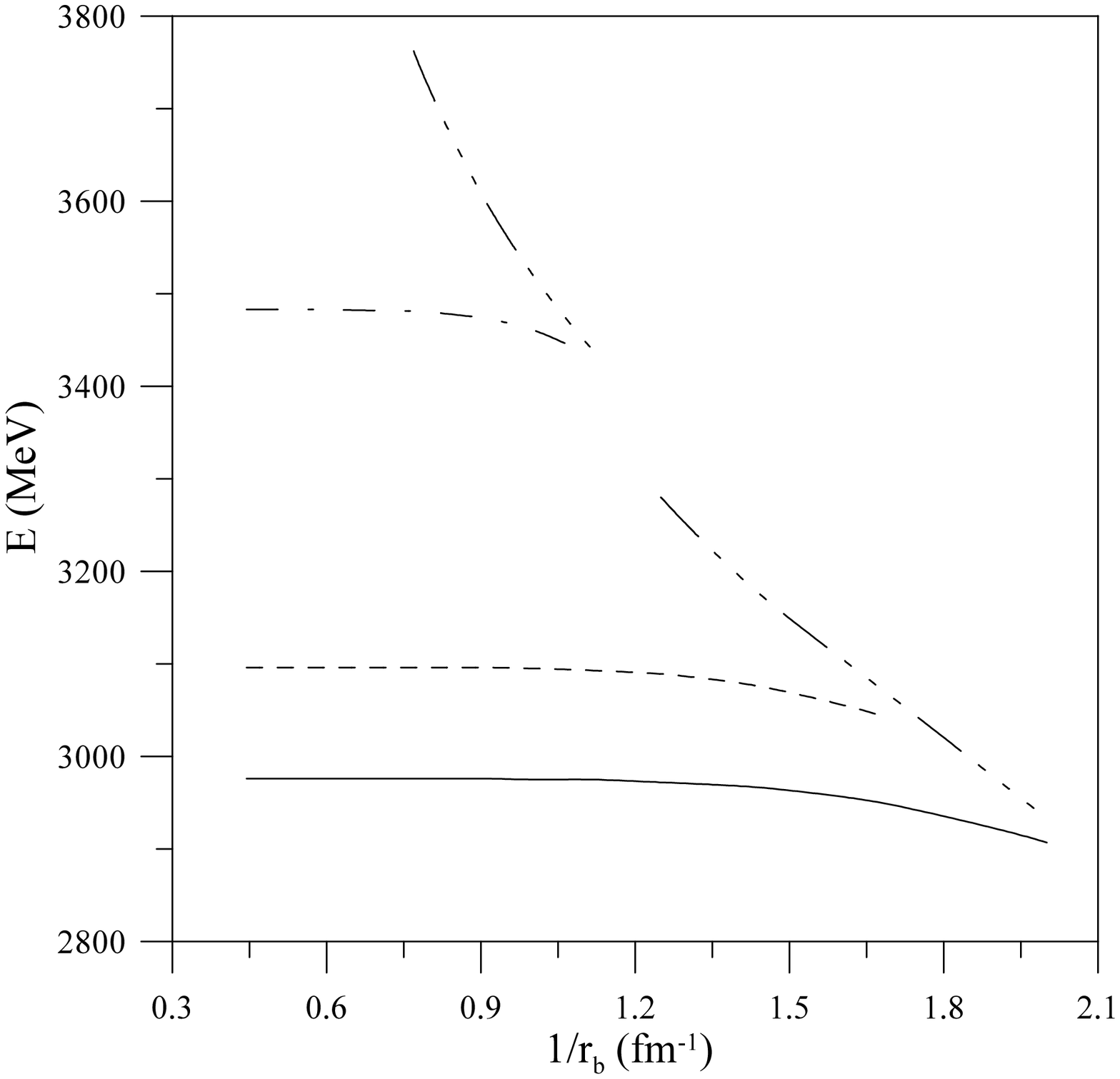,height=70mm}
\caption{$\eta_c$ (solid), $J/\Psi$ (dashed) and $\chi_{cJ}$ (dashed-dotted) energies as a
function of the smooth $\mu$ (upper panel) and sudden $1/r_b$ (lower panel) screening parameters. The
dashed--triple-dotted curve shows the threshold energies of these states.}
\label{fig:E-ccT}
\end{figure}

\begin{figure}[t]
\epsfig{file=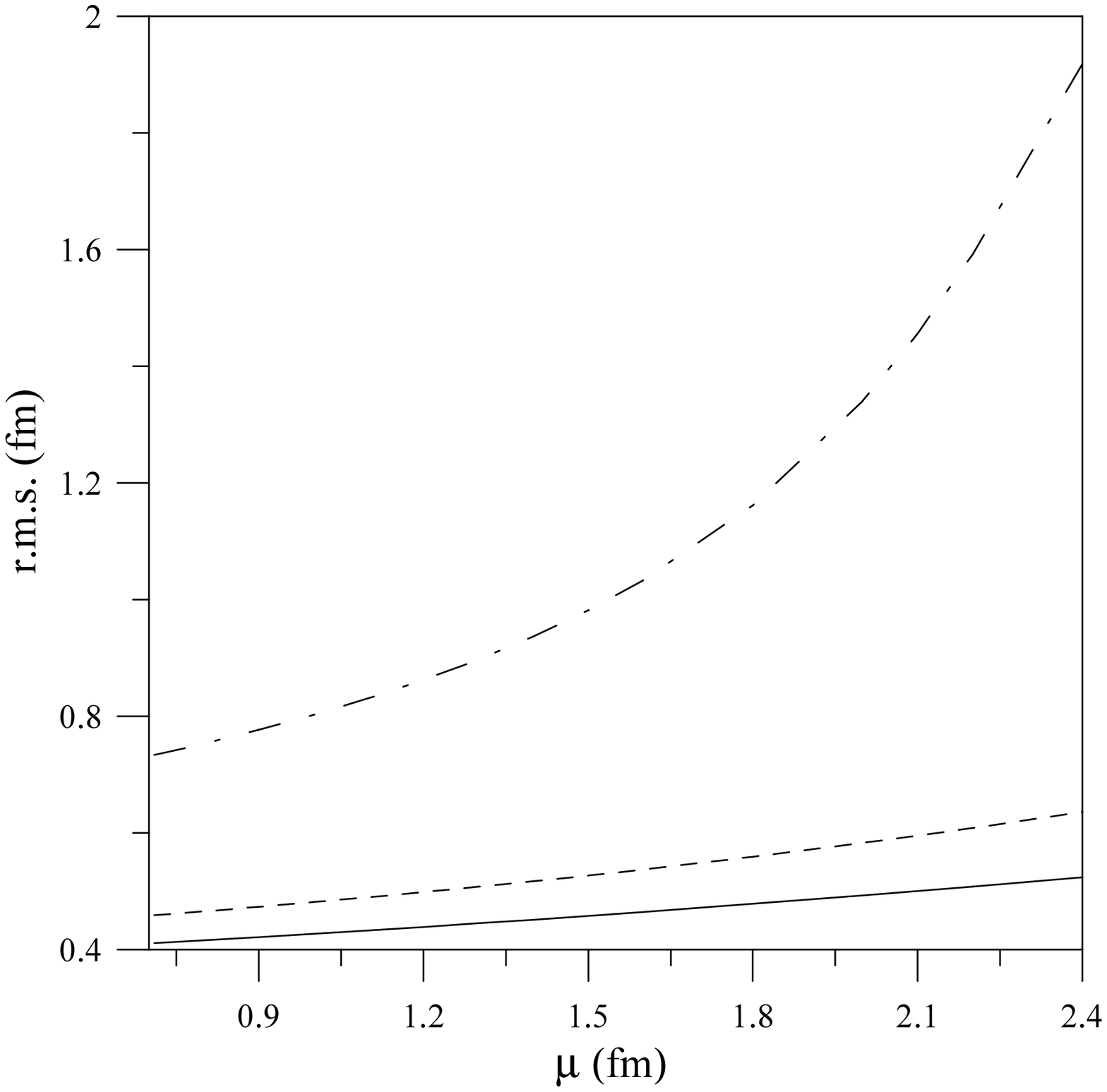,height=70mm}
\epsfig{file=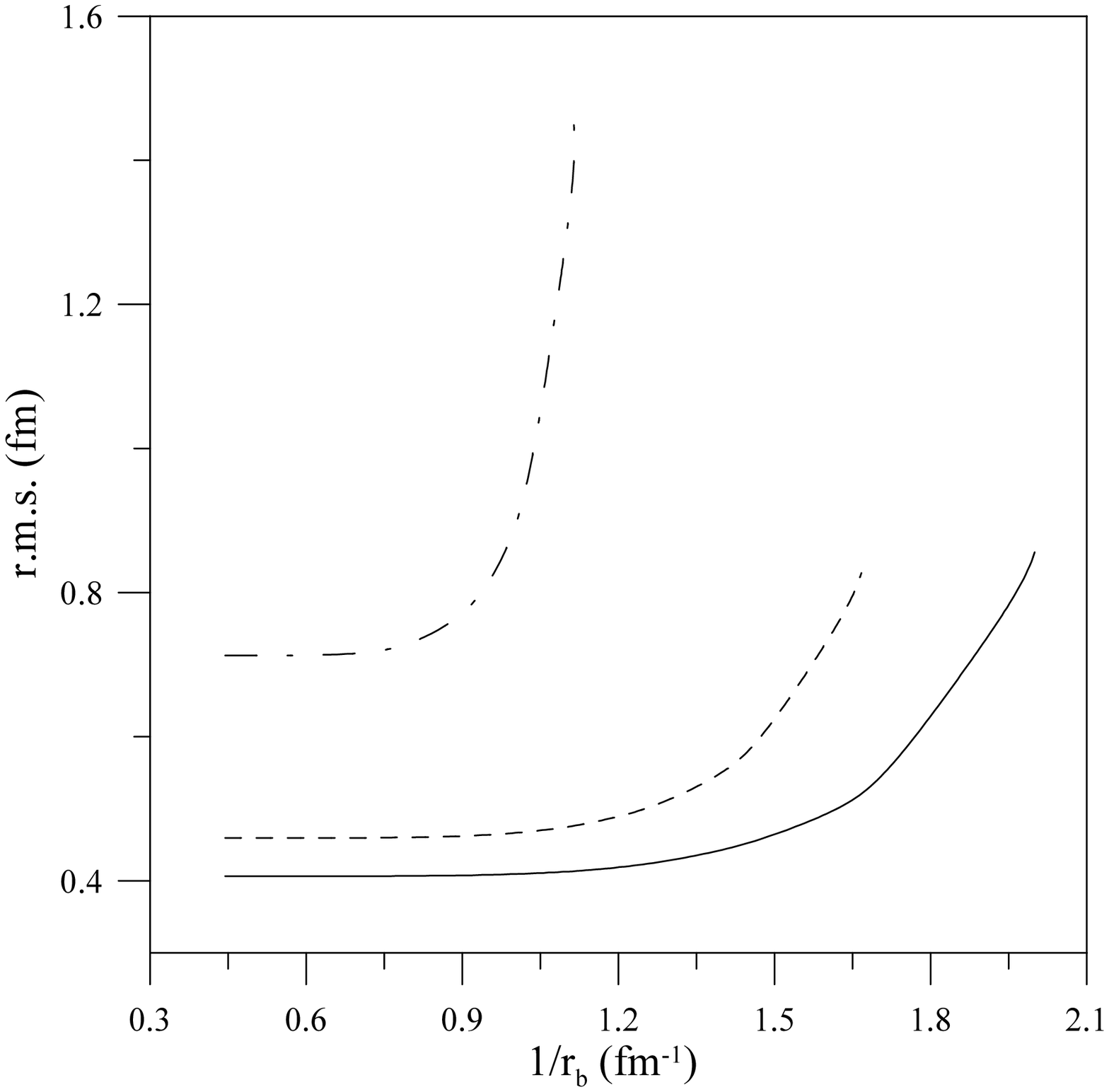,height=70mm}
\caption{$\eta_c$ (solid), $J/\Psi$ (dashed) and $\chi_{cJ}$ (dashed-dotted) r.m.s. as a function
of the smooth $\mu$ (upper panel) and sudden $1/r_b$ (lower panel) screening parameters.}
\label{fig:R-ccT}
\end{figure}

\begin{figure}[t]
\epsfig{file=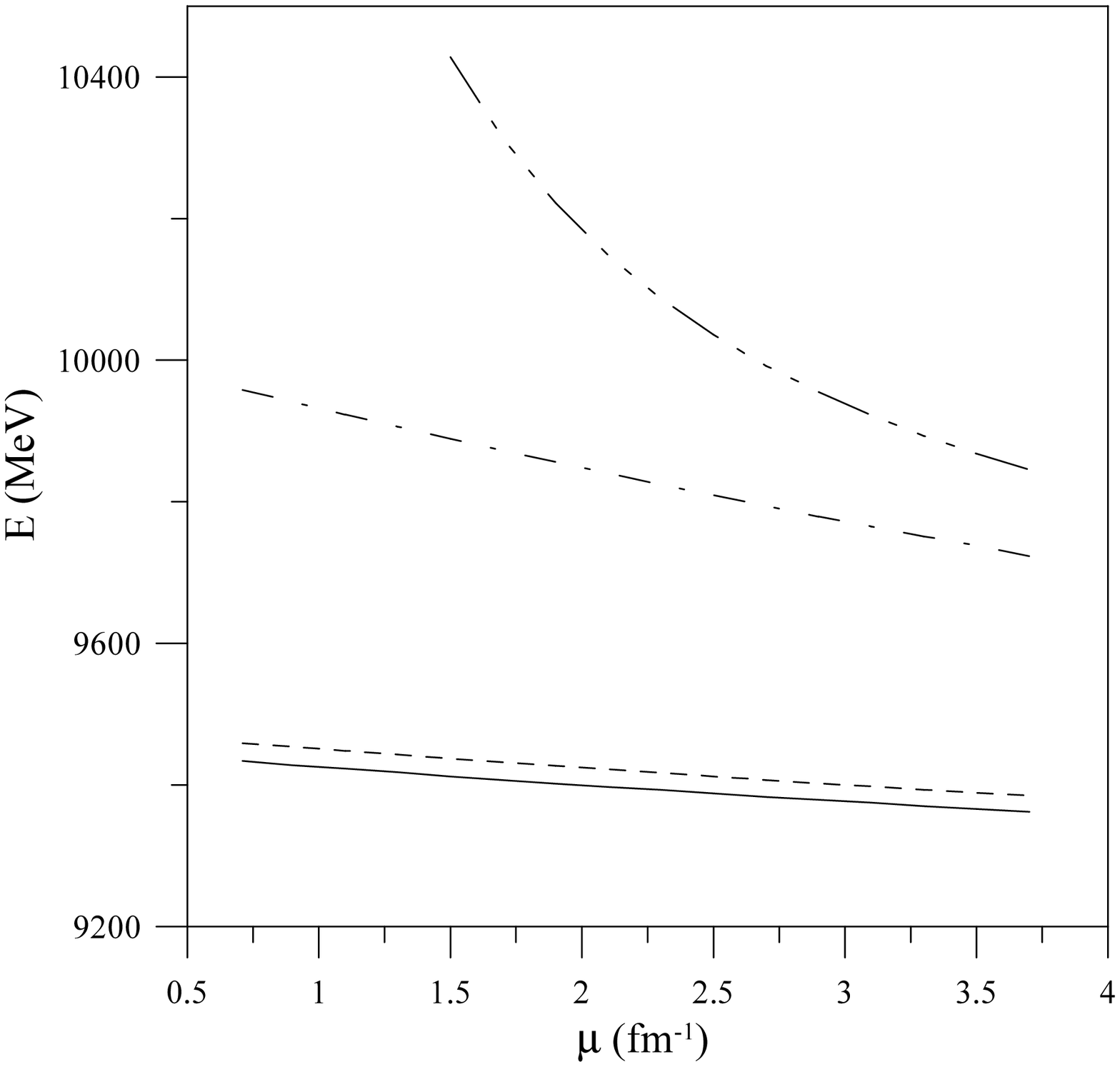,height=70mm}
\epsfig{file=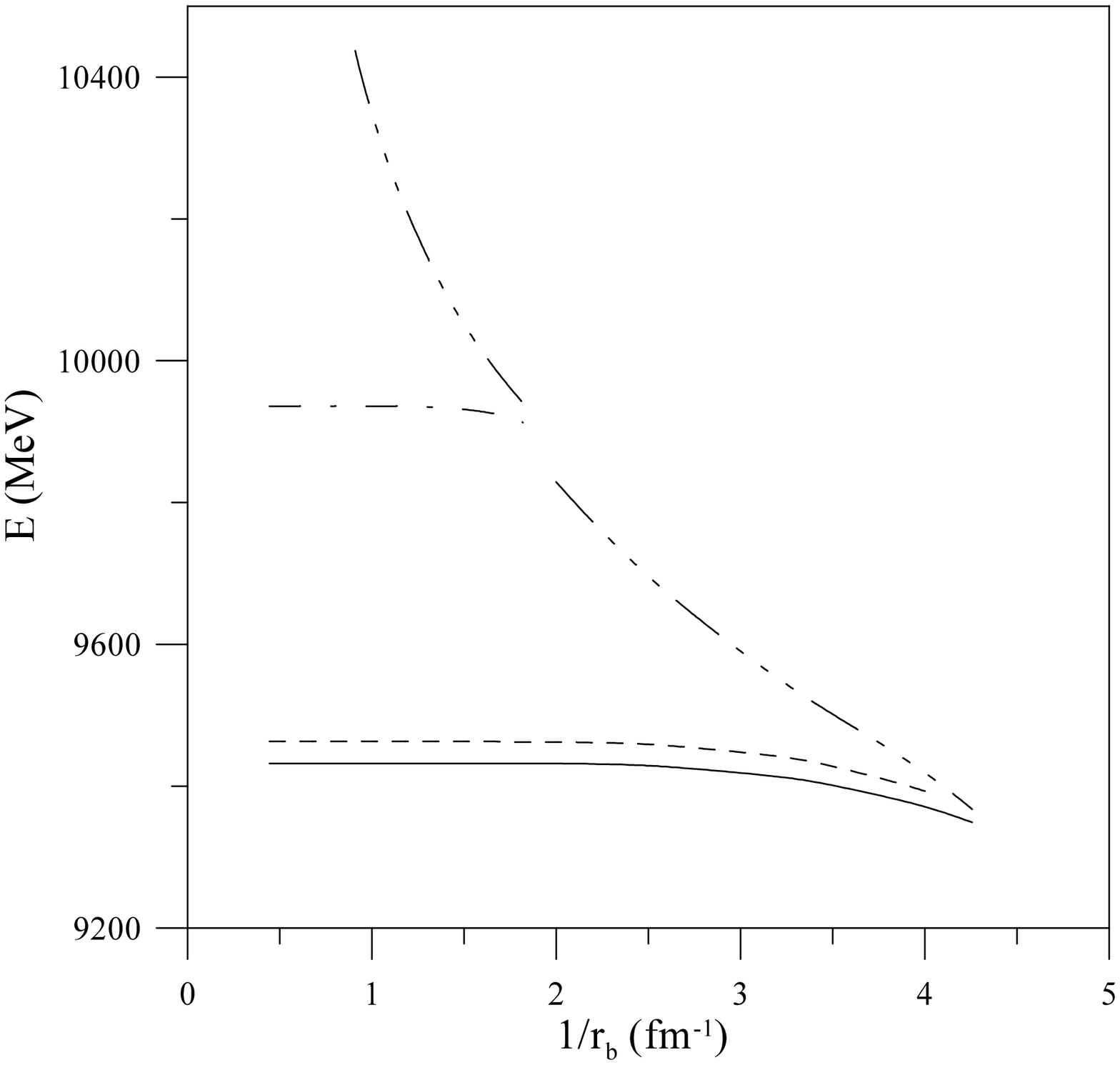,height=70mm}
\caption{$\eta_b$ (solid), $\Upsilon(1S)$ (dashed) and $\chi_{bJ}$ (dashed-dotted) energies as
a function of the smooth $\mu$ (upper panel) and sudden $1/r_b$ (lower panel) screening parameters.
The dashed--triple-dotted curve shows the threshold energies of these states.}
\label{fig:E-bbT}
\end{figure}

\begin{figure}[t]
\epsfig{file=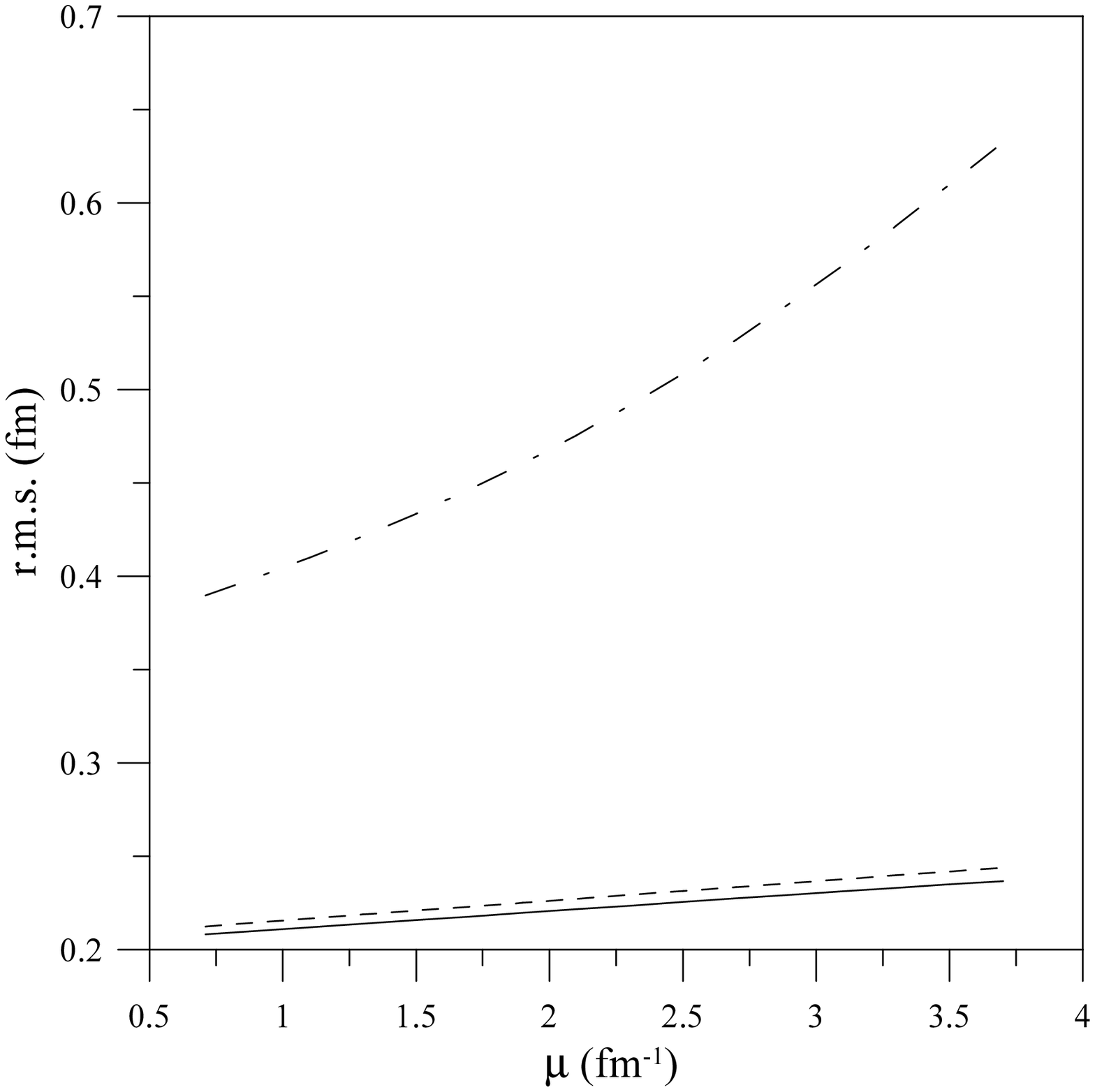,height=70mm}
\epsfig{file=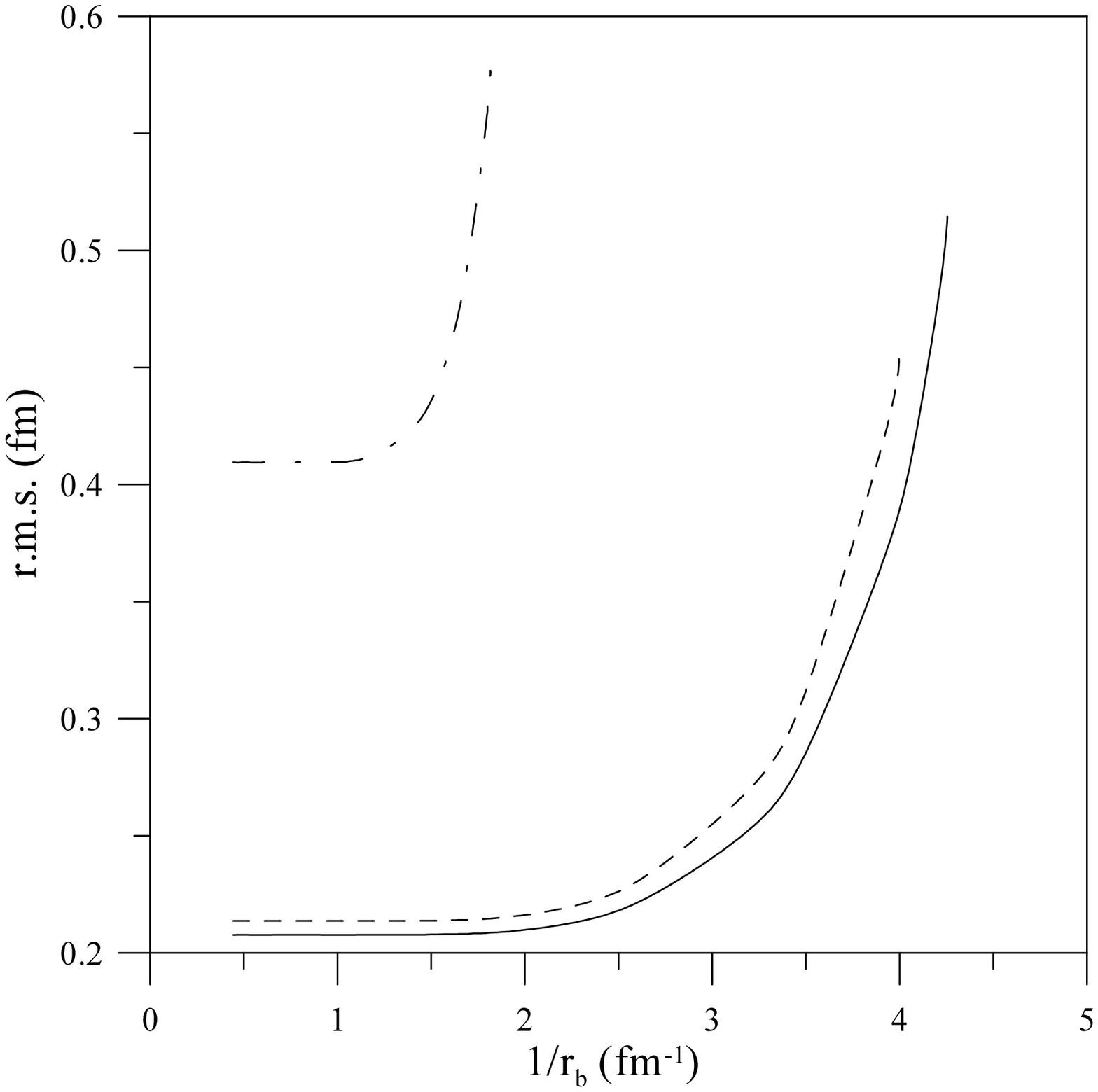,height=70mm}
\caption{$\eta_b$ (solid), $\Upsilon(1S)$ (dash) and $\chi_{bJ}$ (dash-dot) r.m.s. as a
function of the smooth $\mu$ (upper panel) and sudden $1/r_b$ (lower panel) screening parameters.
}
\label{fig:R-bbT}
\end{figure}

Next we keep all parameters fixed and vary the screening parameters $\mu$ and
$1/r_b$ so to mimic a temperature dependence. Here we do not use a specific
model for the temperature dependence of these parameters, this will be
discussed in the following section. Results for the total energy $E$ and r.m.s.
radius $\sqrt{\langle r^2\rangle}$ of the lowest $S$ and $P$ wave states of
charmonium as functions of the smooth ($\mu$) and sudden ($1/r_b$) screening parameters are shown
respectively in Figs.~\ref{fig:E-ccT} and~\ref{fig:R-ccT}. The corresponding
results for bottomonium are shown Figs.~\ref{fig:E-bbT} and \ref{fig:R-bbT}.
In Tables~\ref{tab:ccT} and \ref{tab:bbT} the results corresponding to the sudden
potential are presented together with the threshold energies $E_{th}$ and wavefunctions
at the origin $\phi(0)$.

\begin{table*}[htb]
\caption{Values of the threshold $E_{th}$ and total $E$ energies, in MeV, r.m.s. radius $\sqrt{\langle r^2\rangle}$, in fm,
and wavefunction at the origin $\phi(0)$, in fm$^{-3/2}$, of the lowest charmonium $S$ and $P$ wave states
for different values of the sudden screening parameter $1/r_b$, in fm$^{-1}$.}
\begin{center}
\begin{tabular}{|c|cccc|cccc|cccc|}
\hline
	&\multicolumn{4}{c|}{$\eta_c$}	&\multicolumn{4}{c|}{$J/\Psi$}	 &\multicolumn{4}{c|}{$\chi_{cJ}$} \\
\hline
$1/r_b$	&$E_{th}$	&$E$	 &$\sqrt{\langle r^2\rangle}$   &$\phi(0)$ &$E_{th}$ &$E$ &$\sqrt{\langle r^2\rangle}$
&$\phi(0)$&$E_{th}$ &$E$ &$\sqrt{\langle r^2\rangle}$  &$\phi(0)$ \\
\hline
0.44	& 4523	& 2976.0 & 0.4066 & 13.328 & 4523 & 3096.4 & 0.4596 & 9.486 & 4534 & 3482.6 & 0.7126 & 0 \\
0.48	& 4399	& 2976.0 & 0.4066 & 13.328 & 4399 & 3096.4 & 0.4596 & 9.486 & 4412 & 3482.6 & 0.7126 & 0 \\
0.53	& 4234	& 2976.0 & 0.4066 & 13.328 & 4234 & 3096.4 & 0.4596 & 9.486 & 4250 & 3482.6 & 0.7126 & 0 \\
0.59	& 4067	& 2976.0 & 0.4066 & 13.328 & 4068 & 3096.4 & 0.4596 & 9.486 & 4087 & 3482.6 & 0.7129 & 0 \\
0.67	& 3899	& 2976.0 & 0.4066 & 13.328 & 3900 & 3096.4 & 0.4597 & 9.486 & 3924 & 3482.4 & 0.7145 & 0 \\
0.77	& 3727	& 2975.8 & 0.4067 & 13.328 & 3729 & 3096.3 & 0.4600 & 9.484 & 3762 & 3481.1 & 0.7235 & 0 \\
0.91	& 3552	& 2975.7 & 0.4076 & 13.320 & 3554 & 3096.0 & 0.4624 & 9.468 & 3600 & 3473.1 & 0.7773 & 0 \\
1.00	& 3461	& 2975.4 & 0.4093 & 13.305 & 3465 & 3095.3 & 0.4664 & 9.439 & 3521 & 3461.4 & 0.8892 & 0 \\
1.11	& 3368  & 2974.6 & 0.4132 & 13.262 & 3374 & 3093.5 & 0.4759 & 9.363 & 3443 & 3432.1 & 1.4710 & 0 \\
1.25	& 3271  & 2972.4 & 0.4230 & 13.151 & 3280 & 3088.9 & 0.4998 & 9.170 & \multicolumn{4}{c|}{Melted}\\
1.43	& 3168  & 2966.5 & 0.4483 & 12.856 & 3182 & 3077.0 & 0.5661 & 8.669 & \multicolumn{4}{c|}{Melted}\\
1.67	& 3058  & 2950.6 & 0.5207 & 12.070 & 3078 & 3046.3 & 0.8268 & 7.260 & \multicolumn{4}{c|}{Melted}\\
2.00	& 2933  & 2906.7 & 0.8556 &  9.633 &  \multicolumn{4}{|c|}{Melted}  & \multicolumn{4}{c|}{Melted}\\
2.50    &   \multicolumn{4}{|c|}{Melted}   &  \multicolumn{4}{|c|}{Melted}  & \multicolumn{4}{c|}{Melted}\\
\hline
\end{tabular}
\end{center}
\label{tab:ccT}
\end{table*}

\begin{table*}[htb]
\caption{Values of the threshold $E_{th}$ and total $E$ energies, in MeV, r.m.s. radius $\sqrt{\langle r^2\rangle}$, in fm,
and wavefunction at the origin $\phi(0)$, in fm$^{-3/2}$, of the lowest bottomonium $S$ and $P$ wave states
for different values of the sudden screening parameter $1/r_b$, in fm$^{-1}$.}
\begin{center}
\begin{tabular}{|c|cccc|cccc|cccc|}
\hline
	&\multicolumn{4}{c|}{$\eta_b$}	&\multicolumn{4}{c|}{$\Upsilon(1S)$}	 &\multicolumn{4}{c|}{$\chi_{bJ}$} \\
\hline
$1/r_b$	&$E_{th}$	&$E$	 &$\sqrt{\langle r^2\rangle}$   &$\phi(0)$ &$E_{th}$ &$E$ &$\sqrt{\langle r^2\rangle}$
&$\phi(0)$&$E_{th}$ &$E$ &$\sqrt{\langle r^2\rangle}$  &$\phi(0)$ \\
\hline
0.44	& 11392	& 9432.2 & 0.2077 & 41.530 & 11393 & 9462.8 & 0.2136 & 39.271 & 11396 & 9936.0 & 0.4095 & 0 \\
0.48	& 11269	& 9432.2 & 0.2077 & 41.530 & 11270 & 9462.8 & 0.2136 & 39.271 & 11273 & 9936.0 & 0.4095 & 0 \\
0.53	& 11104	& 9432.2 & 0.2077 & 41.530 & 11104 & 9462.8 & 0.2136 & 39.271 & 11109 & 9936.0 & 0.4095 & 0 \\
0.67	& 10769	& 9432.2 & 0.2077 & 41.530 & 10769 & 9462.8 & 0.2136 & 39.271 & 10776 & 9936.0 & 0.4095 & 0 \\
0.91	& 10423	& 9432.2 & 0.2077 & 41.530 & 10424 & 9462.8 & 0.2136 & 39.271 & 10437 & 9936.0 & 0.4096 & 0 \\
1.11	& 10242	& 9432.2 & 0.2077 & 41.530 & 10242 & 9462.8 & 0.2136 & 39.271 & 10262 & 9935.9 & 0.4105 & 0 \\
1.43	& 10048	& 9432.2 & 0.2078 & 41.528 & 10049 & 9462.8 & 0.2137 & 39.266 & 10082 & 9933.3 & 0.4265 & 0 \\
1.67	&  9942	& 9432.1 & 0.2081 & 41.513 &  9944 & 9462.7 & 0.2141 & 39.250 &  9989 & 9924.6 & 0.4810 & 0 \\
1.82	&  9886	& 9432.0 & 0.2086 & 41.487 &  9888 & 9462.5 & 0.2147 & 39.250 &  9941 & 9912.9 & 0.5859 & 0 \\
2.00	&  9826	& 9431.6 & 0.2098 & 41.423 &  9829 & 9462.1 & 0.2162 & 39.145 & \multicolumn{4}{c|}{Melted}\\
2.50	&  9692	& 9428.6 & 0.2181 & 40.922 &  9696 & 9458.5 & 0.2263 & 38.566 & \multicolumn{4}{c|}{Melted}\\
3.33	&  9521	& 9408.8 & 0.2634 & 38.059 &  9529 & 9436.5 & 0.2829 & 35.318 & \multicolumn{4}{c|}{Melted}\\
4.00	&  9408	& 9370.8 & 0.3893 & 32.075 &  9419 & 9393.3 & 0.4590 & 28.454 & \multicolumn{4}{c|}{Melted}\\
4.26	&  9368	& 9348.6 & 0.5144 & 28.194 &   \multicolumn{4}{|c|}{Melted}   & \multicolumn{4}{c|}{Melted}\\
5.00    &   \multicolumn{4}{|c|}{Melted}   &  \multicolumn{4}{|c|}{Melted}    & \multicolumn{4}{c|}{Melted}\\
\hline
\end{tabular}
\end{center}
\label{tab:bbT}
\end{table*}

The results are quite striking. The observables calculated with
a smooth screening or with sudden screening behave dramatically different
as the corresponding screening parameters are varied. While the observables
calculated with smooth screening vary continuously as $\mu$ is increased,
the observables calculated with sudden screening change very little as $1/r_b$
increases, until a critical value of this parameter is reached when they
change abruptly. Such an abrupt change on observables at a critical value
of $1/r_b$ naturally leads to the interpretation of a phase transition,
the deconfinement transition. Of course, related to this critical value
of $1/r_b$ there should be a critical temperature $T_c$.

\begin{figure}[htb]
\begin{center}
\epsfig{file=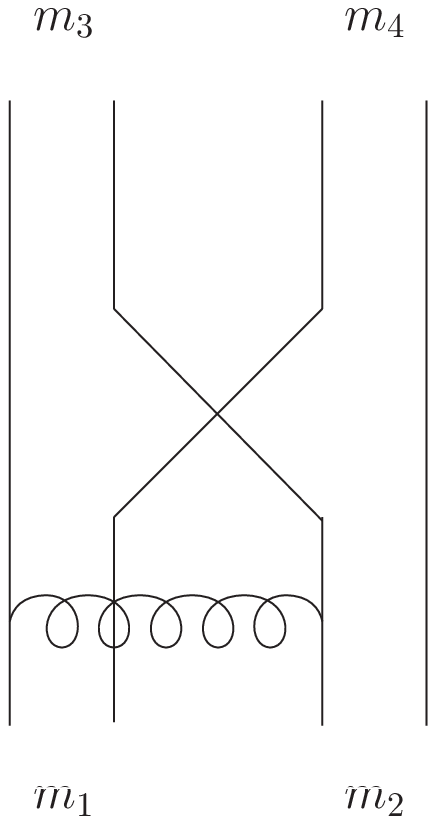,height=70mm}
\end{center}
\caption{Pictorial representation of the constituent interchange mechanism
in meson-meson scattering.}
\label{fig:diag}
\end{figure}

As mentioned in the introduction, an abrupt increase of the size of
the meson wave functions will impact the phenomenology of quarkonia
interactions in medium, in particular for scattering dissociation processes
\cite{{Hadjimichef:1998rx},{Wong:1999zb},{Hilbert:2007hc},{Haidenbauer:2007jq},CBM-Fair}.
In a quark model description, such
dissociation cross sections depend upon the degree of overlap of the wave functions
of the hadrons and so depend crucially on the size of the wave functions in
coordinate space. The diagram shown in Fig.~\ref{fig:diag} illustrates such a
process for the case of a meson-meson process. The figure illustrates a process
in which two mesons $m_1$ and $m_2$ collide and give in general two different
mesons $m_3$ and $m_4$. The basic mechanism here is quark-gluon interchange in
which the two quarks, one from each of the colliding mesons, are interchanged
with each other leading to a final state that is different from the initial
one. A typical and very important example is the process $J/\Psi + \pi
\rightarrow D \bar D^\ast$. It is clear that the constituent interchange will
only happen when there is significant overlap between the wave functions
of two mesons. The same quark-gluon interchange mechanism is also present in
elastic hadron-hadron scattering and has been used in recent
publications to describe the short-range part of several different processes
\cite{Haidenbauer:2007jq,Hadjimichef:2000en}.
As in dissociation processes, the elastic processes are also
very sensitive to the sizes of the wave functions of the colliding
hadrons.

In order to illustrate the effect of an abrupt increase of the size of the
wave functions on a cross section, let us consider quark-gluon interchange
in elastic meson-meson scattering.  To evaluate the cross-section,
we use the quark-Born-diagram method~\cite{QBD1,QBD2}, which provides
a good approximation~\cite{Hadjimichef:1998rx} to a more complete resonating
group method or Born-Oppenheimer calculation~\cite{AV}.
There are several diagrams that contribute to this process, being that shown
in Fig.~\ref{fig:diag} a representative one. The evaluation of such graphs
requires the quark-gluon interaction and the hadron wave functions. The calculation
of these diagrams can be carried out analytically when one uses Gaussian forms
for the hadron wave functions and a contact interaction for the quark-quark
interaction (the wavy line in Fig.~\ref{fig:diag}). To set the notation, let the
internal meson wave functions $\psi(r)$ be given as (for equal quark masses)
\begin{equation}
\psi(r) = \left(\frac{1}{2\pi}\right)^{3/2}\left(\frac{\pi}{\lambda}\right)^{3/4} e^{-r^2/8\lambda},
\end{equation}
so that the r.m.s. radius of the meson is given by
$\langle r^2 \rangle = 6 \, \lambda$. In general, as the c.m. total
energy of the meson-meson system $s$ is increased, most of the Born elastic cross
section decreases very rapidly and leaves a constant cross section at high energies,
when $s \gg 1/\lambda$ -- see discussions in Ref.~\cite{QBD1}.
Specifically, the behavior of the cross section at large $s$ is given by
\begin{equation}
\lim_{s \gg 1/\lambda} \sigma \sim \frac{1}{\lambda} \sim \frac{1}{\langle r^2 \rangle} .
\label{diff_x}
\end{equation}
On the other hand, at low momentum transfers $t$ such that $t/s$ is small, the
differential cross section behaves as
\begin{equation}
\lim_{t/s \ll 1} \frac{d\sigma}{dt} \sim \frac{1}{\lambda} e^{\lambda t} .
\label{tot_x}
\end{equation}
Although these results are for the case when all meson wave functions have the
same size $\lambda$, they are of general validity. The final expression would
be more complicated when meson wave functions of different sizes are used,
but the behavior of the above cross section would be qualitatively the same,
in that they decrease when the size of any of the meson wave functions increases.

From Eqs.~(\ref{diff_x}) and (\ref{tot_x}) it is evident the role played
by the size of the meson wave functions. When the size of the wave functions
increase abruptly as the confining string breaks, the high energy cross sections
that involve constituent interchange will change abruptly. As the critical point
is crossed, the model discussed here predicts that the cross sections involving
charmonium can decrease by a factor of five before the hadrons melt.
For bottomonium the decrease is a factor of two.

The physical reason for the decrease of the cross sections is of course due to
the fact that the hadron wave functions flatten out as their size parameters
increase, since they are normalized. As the hadron wave functions flatten out,
the overlap of the wave functions of the colliding hadrons is less
significative and as such the probability of constituent interchange is
smaller. This effect is similar to the decrease of the electroweak decay constants
as the size parameter increases. Since the decay constants are proportional
to the wave function at the origin $\phi(0)$, they will decrease because
$\phi(0)$ decreases as the extension of the wave function increases
-- see Tables~\ref{tab:ccT} and~\ref{tab:bbT}.

\section{A model for the temperature dependence of screening parameters}
\label{sec:temperature}

A phenomenological relation between the sreening parameters, $\mu$ in
Eq.~(\ref{Vsmooth}) or $r_b$ in Eq.~(\ref{Vsudden}),
with temperature can be made within the
approach of Ref.~\cite{Gonzalez:2003gx}, where the screening parameter is connected
to an effective gluon mass scale in the infrared. Although a particular framework will
be used to substantiate this, we trust that it might be of general validity. Suppose
initially that the heavy-quark potential in coordinate space is defined as the Fourier
transform of the static one-gluon-exchange with a running coupling constant
$\alpha_s(Q^2)$. For the sake of argument, let us suppose~\cite{Gonzalez:2003gx} that
$\alpha_{s}(Q^{2})$ is given by the form derived in QCD by Cornwall long
ago~\cite{Cornwall}, namely
\begin{equation}
\alpha _{s}(Q^{2})=\frac{4\pi }{\beta _{0}\ln \left[
(Q^{2}+4M_{g}^{2}(Q^{2}))/\Lambda ^{2}\right] },
\label{RC}
\end{equation}
where $\beta _{0}=(33-2n_{f})/3$, $n_{f}$ is the number of quark flavors with
masses much smaller than $Q$, $\Lambda \sim 300$~MeV is the QCD scale parameter, and
$M_{g}(Q^2)$ is an effective running gluon mass given by
\begin{equation}
M_{g}^{2}(Q^{2})=m_{g}^{2}\,\,\left( \frac{\ln \left[ (Q^{2}+4m_{g}^{2})/%
\Lambda ^{2}\right] }{\ln \left( 4m_{g}^{2}/\Lambda ^{2}\right)}
\right)^{-12/11} ,
\label{GM}
\end{equation}
with $m_{g}$ being a constant mass scale and is responsible for the existence
of confinement. One has that $\alpha_s(Q^2)$ runs from 0 in the ultraviolet
asymptotic freedom limit $Q^{2}\rightarrow \infty$, to $\alpha(0) = 4\pi/\beta_0
\ln (4m^2_g/\Lambda^2)$ in the deep infrared limit of $Q^2 \rightarrow 0$.
Now, for $ M^2_g(Q^2) \sim \Lambda^2$ one obtains a linearly-rising potential in
coordinate space, since
in this limit $\alpha_s(Q^2) \rightarrow 1/Q^2$. Strictly speaking this occurs for
the precise value $M^2(Q^2) = \Lambda^2/4$, otherwise the potential is not exactly
linear, but still absolutely confining. Therefore, the potential one gets is of
Coulomb type at short distances (modified by asymptotic freedom) and
confining at long distances. Obviously, both short- and long-distance
components of this potential do not contain the effects of screening. However,
screening can be modeled by modifying the potential so that its long distance
component saturates at a distance $1/\mu$, where $\mu$ is a screening mass.
Screening goes away for $\mu =0$ and the purely confining potential with
$M_g \sim \Lambda$ is recovered. In view of the interplay between $\mu$
and $M_g$ it is natural to propose~\cite{Gonzalez:2003gx},
\begin{equation}
\mu =\Lambda -M_{g}\, . \label{MU}
\end{equation}
These effects were parametrized in
exploratory lattice studies by a screened funnel potential~\cite{Bor80}.
Although this para\-metrization does not reproduce the rapid
turnover around 1 fm from linearly rising to flat potential
suggested by modern lattice results~\cite{C}, we will follow
it for the sake of simplicity. For a meson the
above reasoning would give rise to a confining
static potential of the form
\begin{equation}
V_{conf}(r)=\frac{\sigma }{\mu }-\sigma r%
\frac{e^{-\mu r}}{\mu r}=\frac{\sigma }{\mu }-\sigma r%
\left[ \frac{e^{-(\Lambda -M_{g}(Q_0^{2}))r}}{(\Lambda -M_{g}(Q_0^{2}))r}\right]
\,  \label{VCONF}
\end{equation}
where $Q_0$ is the running scale of $\mu$. In this equation
the relation between $\mu$ and $M_g$ has been made explicit.
This identification establishes a deep connection between the saturation of
the coupling constant and the interquark pair creation mechanism both
effects governed by $M_{g}(Q_0^{2})$. Therefore $\mu $ runs with $Q_0^{2}$ so
that $0\simeq \mu (Q_0^{2}=0)\leq \mu (Q_0^{2})\leq \mu (Q_0^{2}\rightarrow \infty
)\simeq$ 1.52 fm$^{-1}$. In this approach, finite temperature effects
can be introduced in the potential by making the effective running gluon mass
temperature dependent, in such a way that for large temperatures
confinement would disappear. This would imply $M_g^2 (Q_0^2,T)$
must decrease with temperature for a fixed value of $Q_0^2$.
Thus, the temperature dependence translates into the screening
parameter through Eq.~(\ref{MU}), that could now be
written as
\begin{equation}
\mu (Q_0^2,T)=\Lambda -M_{g}(Q_0^2,T)\,  \label{MUQ}
\end{equation}
giving rise to a screening parameter $\mu$ increasing with
temperature. Assuming that the scale dependence of $M_g$ at finite
temperature is still similar to Eq.~(\ref{GM})
and making use of Eq.~(\ref{MUQ}), one can obtain the temperature dependence of
the screening parameter. Finally, making use of the
typical momentum of charmonium that
could be assimilated to its reduced mass,
one can then obtain $\mu(Q^2_c,T) = \mu_c(T)$ appropriate for
charmonium. Similar conclusions were obtained in
Refs.~\cite{Satz:2005hx,Kar88}. In particular, 
Ref.~\cite{Kar88} assumed a linear dependence of $\mu$ on $T$
as obtained in first lattice estimates of screening
in high temperature SU(N) gauge theory.

The above reasoning could be repeated for a rapid tur\-nover transition
potential as that of Eq.~(\ref{Vsudden}) obtaining the 
same conclusions. Admittedly this is a very
crude way to obtain the temperature dependence of the gluon mass scale, it does
seem to make physical sense, in view of the expectation that confinement
goes away at sufficiently high temperatures.

\section{Conclusions and outlook}
\label{sec:concl}

To summarize, we have performed a detailed quark model calculation of the $b\overline{b}$ and
$c\overline{c}$ sectors at zero and finite temperature comparing results
obtained using smooth and sudden string breaking potentials.
The scale- and tempera\-ture-dependence of the screening parameters $\mu$ and $1/r_b$ has been
discussed. Such a dependence has been motivated by lattice QCD simulations at finite temperature.
The properties of quarkonia close to the critical deconfining temperature depend strongly
on the choice made for the screening, sudden or smooth. When a sudden
string breaking potential is preferred, mesons are unaffected
by the temperature increase up to the vicinity of the critical temperature $T_c$.
Once this temperature is exceeded the radii increases suddenly and the energies and wave functions
close to the origin drop.
As opposed to this, when a smooth screening potential is considered meson properties respond to
any modification in the temperature in a continuous way, and therefore reacting even to small changes
of the temperature.
Such a different behavior will modify drastically the phenomenology of quarkonia interactions in medium,
in particular for scattering dissociation processes. Heavy-ion collision experiments,
like FAIR at GSI, are ideally suited to discriminate between both possibilities
and therefore to provide an important ingredient in order to clearly specify the {\it long range} ($\approx$2 fm)
structure of confinement.

\begin{acknowledgement}
This work has been partially funded by the Spanish Ministerio de
Educaci\'on y Ciencia and EU FEDER under Contracts No. FPA2007-65748
and PCI2005-A7-0312,
by Junta de Castilla y Le\'{o}n under Contract No. SA016A17, and by the
Spanish Consolider-Ingenio 2010 Program CPAN (CSD2007-00042). Partial financial
support by the Brazilian agencies CNPq and FAPESP is also acknowledged.
\end{acknowledgement}

\end{document}